\documentclass[slac_one]{revtex4}
\usepackage{graphicx}
\usepackage{fancyhdr}
\begin{document}

\title{Exclusive processes at HERA} 

%

\author{Robert Ciesielski (on behalf of the H1 and ZEUS collaborations)}
\affiliation{Deutsches Elektronen-Synchrotron DESY, Notkestr. 85, 22607 Hamburg, Germany}

\begin{abstract}
An increased precision of HERA data allows studies of exclusive processes in their transition from soft to hard interactions as well as in the pQCD domain. The recent measurements of vector meson production and Deeply Virtual Compton Scattering (DVCS) performed by the H1 and ZEUS experiments are reported and compared to pQCD expectations.
 
\end{abstract}

\maketitle

\thispagestyle{fancy}


\section{INTRODUCTION} 

\begin{figure}
\includegraphics[width=0.27\textwidth]{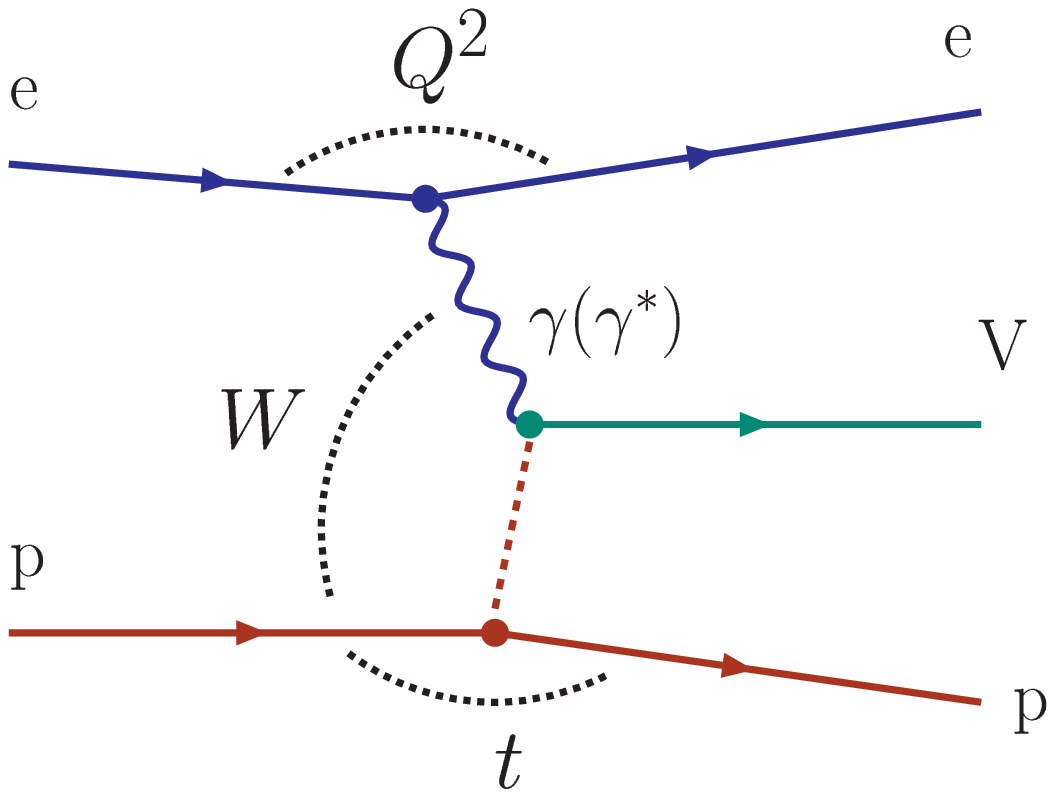}
\hspace{1cm}
\includegraphics[width=0.25\textwidth]{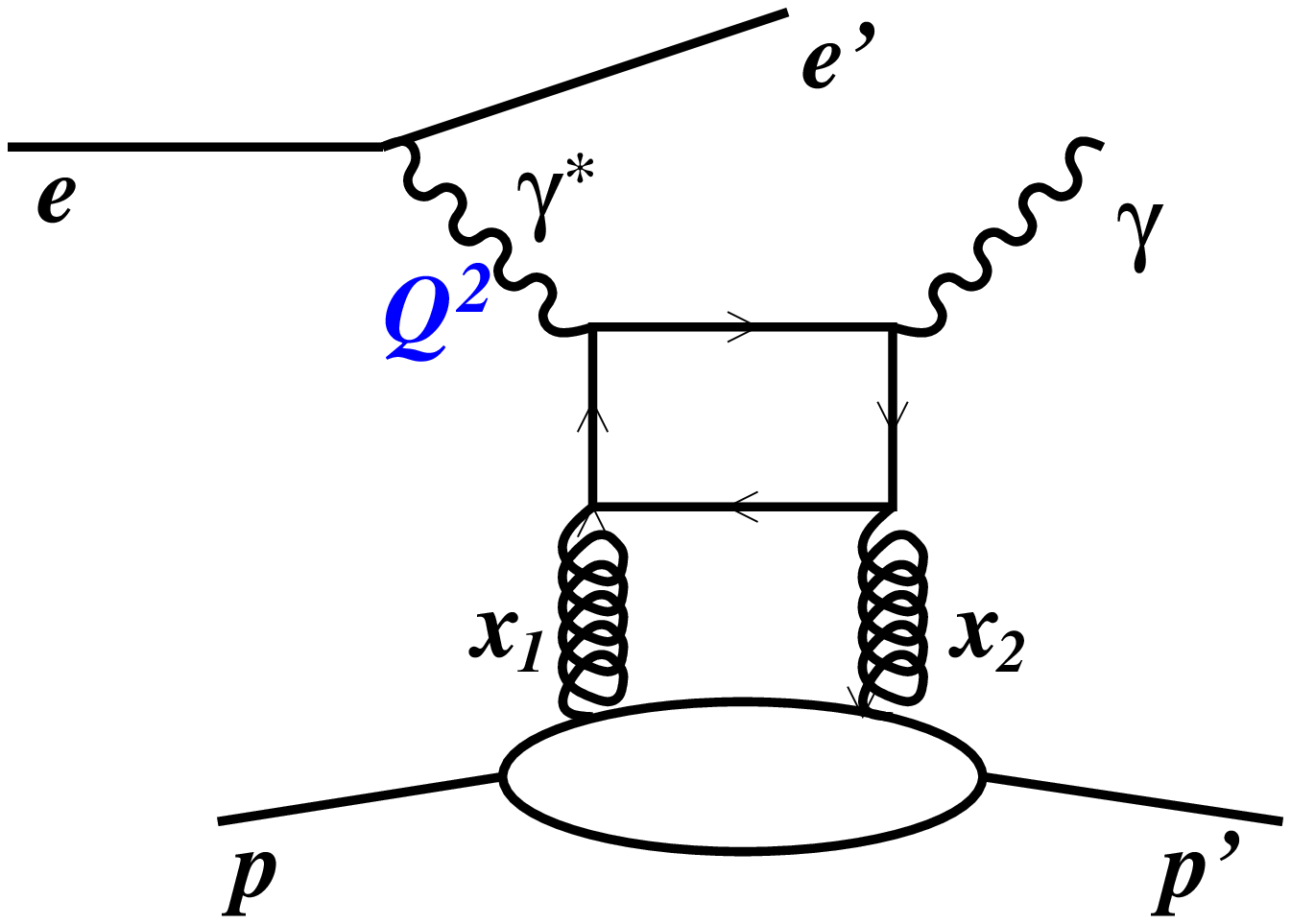}
\caption{A schematic diagram of exclusive vector meson production at HERA, $\gamma^{(*)}p \rightarrow V~p$ (left), and the next-to-leading order Feynman diagram for the DVCS process, $\gamma^{*}p \rightarrow \gamma ~p$ (right).}
\label{fig1}
\end{figure}

Exclusive processes, such as diffractive production of vector mesons, $\gamma^{(*)}p \rightarrow V~p$ with $V=\rho,~\omega,~\phi,~J/\psi,~\psi ', \Upsilon$, and Deeply Virtual Compton Scattering (DVCS), $\gamma^{*}p \rightarrow \gamma~p$, shown in Figure \ref{fig1}, have been extensively studied at HERA. The variables that describe the exclusive process are (Figure \ref{fig1}): the mass of the vector meson, $M$, the photon virtuality, $Q^2$, the four-momentum transfer squared at the proton vertex, $|t|$, and the centre-of-mass energy of the $\gamma p$ system, $W$. The latter is related to the Bjorken scaling variable, $x$, as $W^2 \propto 1/x$. The HERA data span an exceptionally wide range of kinematic variables and thus they can be used to study diffractive interactions in the region where the transition from soft to hard processes occurs, with the hard scale provided by $Q^2$, $M$ or $|t|$. The data can be used to investigate diffraction in terms of pQCD and to test the non-perturbative quantities, such as the generalised or skewed parton densities (GPDs) which describe the correlation between partons in the proton.


In the absence of a hard scale the vector meson production process can be described within the Regge formalism and proceeds by the exchange of the soft Pomeron trajectory. In analogy to low energy hadron-hadron interactions, the slow rise of the cross section with energy ($W^{\delta}$ with $\delta \approx 0.2$), the slow rise of the $b$ slope for the exponential $|t|$ dependence with $W$ and the s-channel helicity conservation (SCHC) are expected. 

In QCD models, at high energies vector meson production factorises into three steps: $(i)$ the emitted photon fluctuates into a $q\bar{q}$ pair (a colour dipole), $(ii)$ the colour dipole interacts with the proton by the exchange of a two-gluon system (or a gluon ladder) in a colour singlet state, $(iii)$ the colour dipole forms the vector meson. Such a description is valid whenever the transverse size of the $q\bar{q}$ pair is small, eg. at high values of $Q^2$ ($Q^2 > \mathcal{O}(10$ GeV$^2)$) or for heavy vector mesons. The process is sensitive to the square of the gluon distribution in the proton and the cross section is expected to rise steeply with energy ($W^{\delta}$ with $\delta \ge 0.7$) in accordance with gluons rising steeply at low-$x$. Furthermore, a universal value of the $b$ slope and violation of the SCHC are expected.

In contrast to vector meson production, the DVCS process with a real photon in the final state is free from the uncertainty of modelling the vector meson wave function. 
This allows access to the skewed parton distribution functions with a better precision. An additional advantage of DVCS is that it interferes with the Bethe-Heitler process (QED $ep \rightarrow ep\gamma$) so that the real part of the amplitude can be measured through the interference term \cite{dvcstheo1}.

This paper reviews the recent high precision measurements of light vector mesons ($\rho$, $\phi$) and DVCS performed by the H1 and ZEUS collaborations.

\section{RESULTS}

\subsection{$W$ dependence of the cross section}


\begin{figure}[t]
\centering
  \includegraphics[width=0.41\columnwidth]{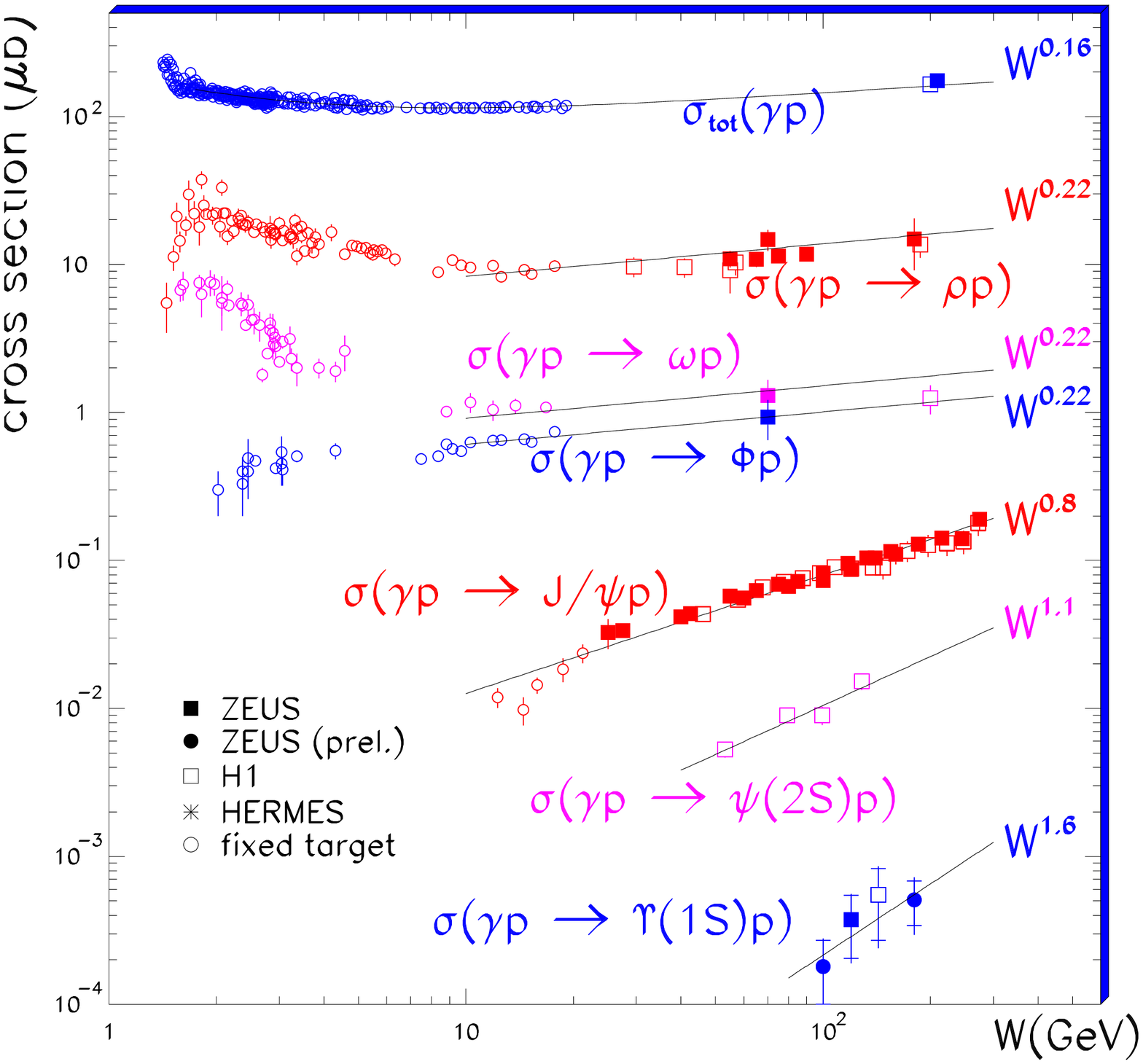}
  \includegraphics[width=0.42\columnwidth]{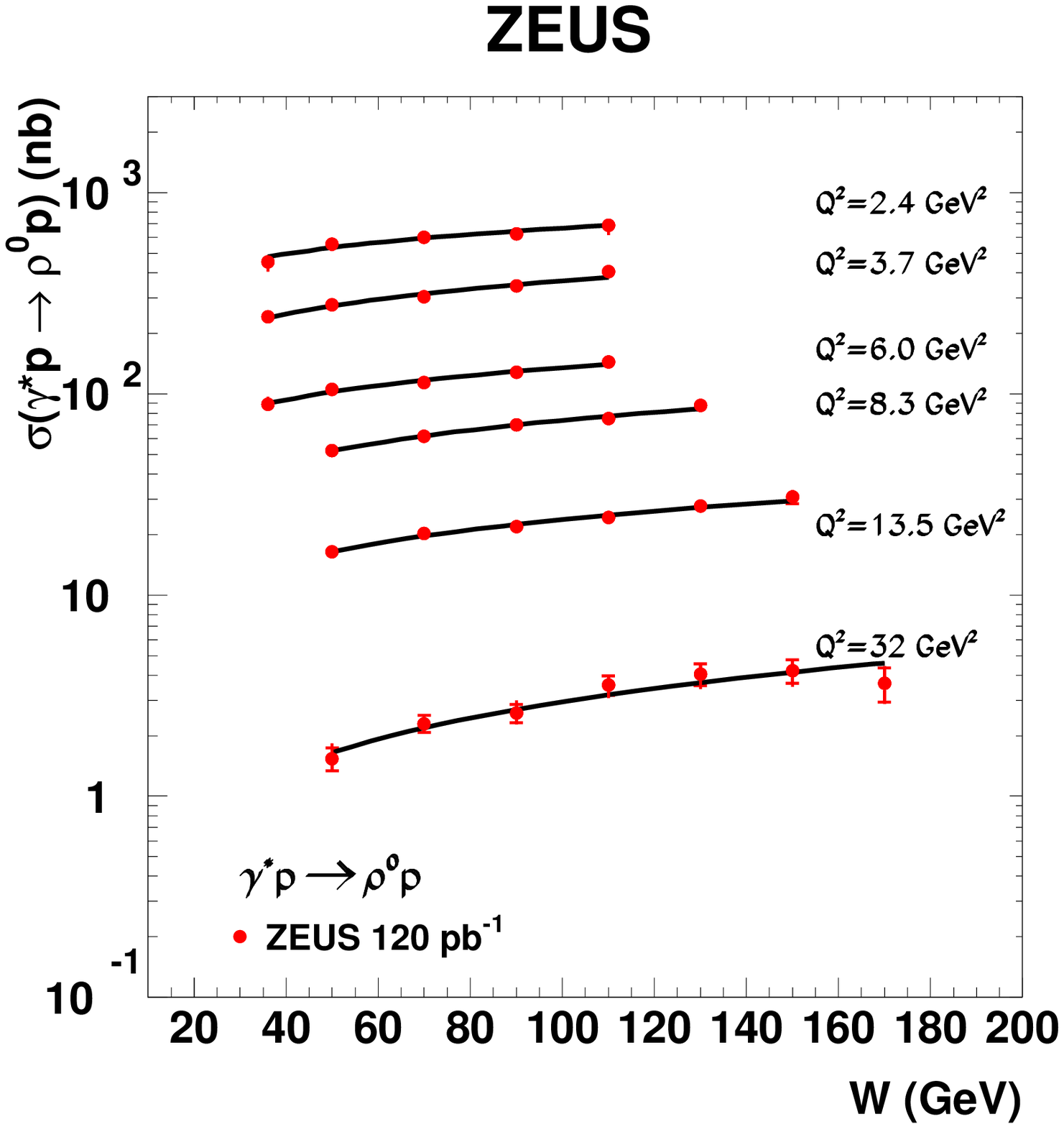}
\caption{A compilation of the cross sections as a function of $W$ for the exclusive photoproduction of various vector mesons, $\gamma p \rightarrow V p$. The total $\gamma p$ cross section and the results from fixed target measurements are also shown. Lines represent the parametrisation with the form $W^{\delta}$ (left). The cross section for the exclusive $\rho$ production, $\gamma^{*} p \rightarrow \rho p$, measured as a function of $W$ at different values of $Q^2$. Lines represent the fits with the function $W^{\delta}$ (right).} 
\label{fig2}
\end{figure}

The $W$ dependence of the cross section for the exclusive photoproduction ($Q^2=0$) of various vector mesons is shown in Figure \ref{fig2}, together with the total $\gamma p$ cross section. At higher energies and for light vector mesons ($\rho, ~\omega, ~\phi$) the cross sections rise slowly with energy, as expected from the soft Pomeron exchange. The values of $\delta \approx 0.22$ ($W^{\delta}$) are similar to that which describes the total cross section. The cross section for the heavy vector mesons, $J/\psi$, $\psi(2S)$ and $\Upsilon(1S)$, exhibits a steeper rise with energy ($\delta \ge 0.7$), which is a signature of the hard process. These observations confirm that the production mechanism changes with the mass of the vector meson. 

The change of the $W$ dependence of the cross section can also be seen for the light vector mesons, $\rho$ and $\phi$ \cite{rhozeus,rhophih1}, as $Q^2$ increases. Figure \ref{fig2} shows the fits with the form $W^{\delta}$ to the $\rho$ data, measured by the ZEUS experiment at six values of $Q^2$ \cite{rhozeus}. The values of $\delta$ are shown as a function of $Q^2+M^2$ in Figure \ref{fig3} (left), together with results for other mesons and DVCS. A universal rise of $\delta$ can be observed, implying that exclusive processes undergo a transition from the soft to the hard regime as $Q^2+M^2$ increases. At higher values of $Q^2+M^2$ the behaviour of $\delta$ is driven by the behaviour of the gluon distribution in the proton. In particular, the recent results for the $\Upsilon$ photoproduction are consistent with the steeper rise of the cross section with $W$ ($\delta \approx 1.6$)
 \cite{upsilonzeus}.

\subsection{$|t|$ dependence of the cross section}

\begin{figure}[t]
\centering
  \includegraphics[width=0.39\columnwidth]{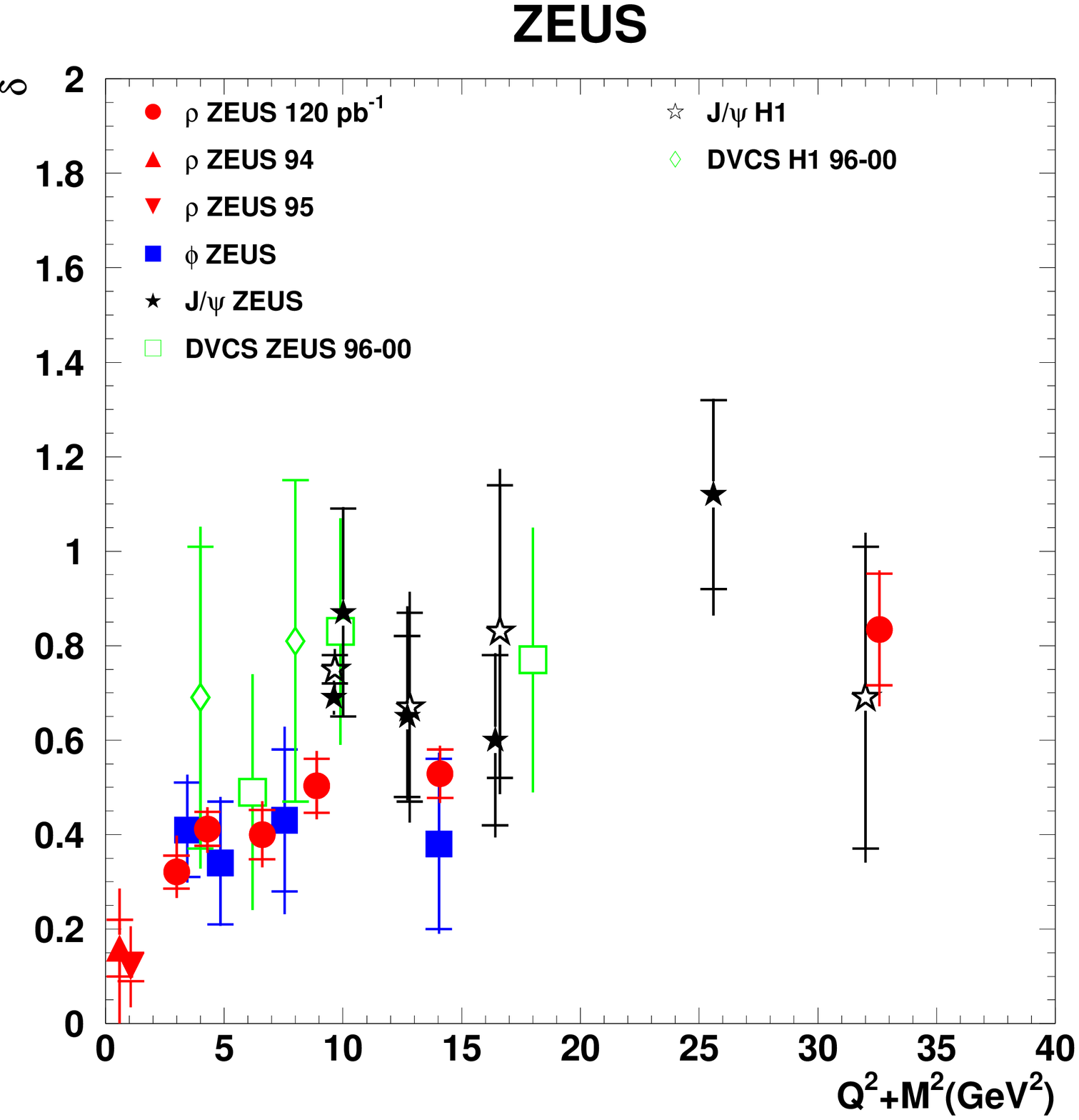}
  \includegraphics[width=0.39\columnwidth]{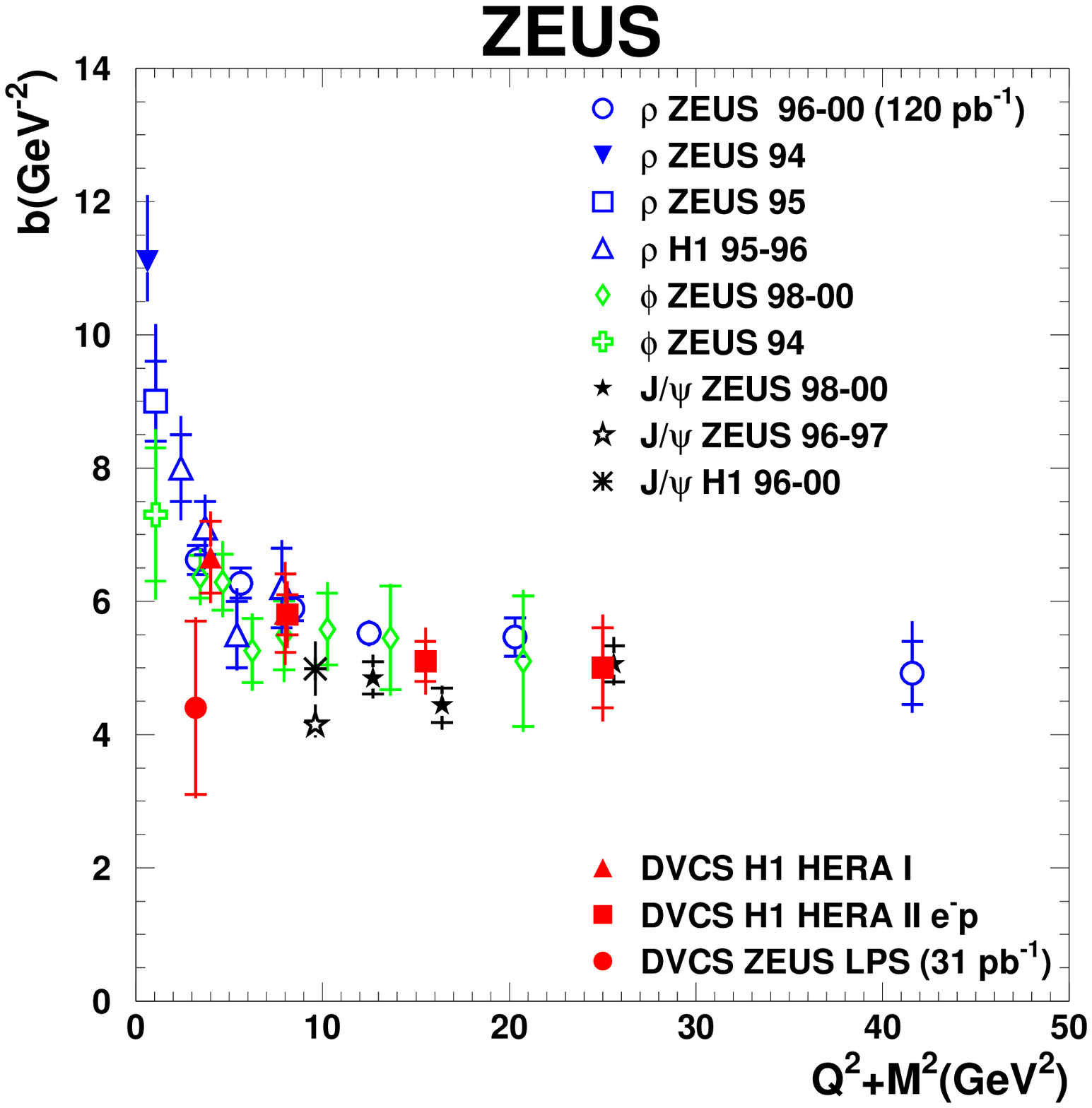}
\caption{The parameter $\delta$ describing the $W$ dependence ($W^{\delta}$) of the cross section (left) and the parameter $b$ describing the $|t|$ dependence ($d\sigma/d|t| \sim e^{-b|t|}$) of the cross section (right) as a function of $Q^2+M^2$ for different vector mesons and DVCS.} 
\label{fig3}
\end{figure}

The differential cross sections as a function of $|t|$ can be described by an exponential function, $\exp{(-b|t|)}$, where the $b$ slope is a measure of the transverse size of the interaction region. The parameters $b$ extracted from the fits to the data for all vector mesons and DVCS \cite{dvcszeus} are shown in Figure \ref{fig3} (right), as a function of $Q^2+M^2$. A universal behaviour of $b$ is observed. It decreases as $Q^2+M^2$ increases due to the shrinking of the dipole size with $Q^2+M^2$. At higher values of $Q^2+M^2$ ($\mathcal{O}(10$ GeV$^2)$), it levels off at $b \approx 5$ GeV$^{-2}$, which corresponds to the transverse size of the gluon cloud of the proton. These results support the pQCD expectations for an universal $b$ slope and point-like character of the interaction in the presence of a hard scale.
	
\subsection{Angular distributions}

The angular distributions of the vector meson decay products depend on three helicity angles and 15 combinations of spin density matrix elements (SDMEs) \cite{schilling}. Their measurement provides information on the spin state of vector mesons and spin dependent properties of the process, for example the SCHC hypothesis that the vector meson retains the helicity of the incoming photon ($\gamma^{*}_L\rightarrow V_L$, $\gamma^{*}_T\rightarrow V_T$), observed in soft processes. Both experiments measured the values of SDMEs in bins of $Q^2$ and in bins of $|t|$ for the $\rho$ and $\phi$ mesons \cite{rhozeus,rhophih1} and found that the SDME $r^5_{00}$ violates the SCHC hypothesis. This observation is in agreement with the expectations from pQCD according to which the orbital momentum of exchanged gluons can contribute to the helicity flip between the incoming photon and the outgoing vector meson.

Furthermore, under assumption of SCHC, vector meson production offers a unique opportunity to extract the ratio $R=\sigma_L/\sigma_T$ and separate the cross sections induced by longitudinally and transversely polarised photons. The ratio $R$ was measured in bins of $Q^2$, $W$ and $|t|$ for the $\rho$ and $\phi$ mesons \cite{rhozeus,rhophih1}. It increases with $Q^2$, implying that $\sigma_L$ becomes dominant at higher $Q^2$. $R$ is also independent of $W$ and $|t|$, but decreases with the mass of the $\rho$ meson.

\subsection{Comparison to theory}

\begin{figure}[t]
\centering
  \includegraphics[width=0.37\columnwidth]{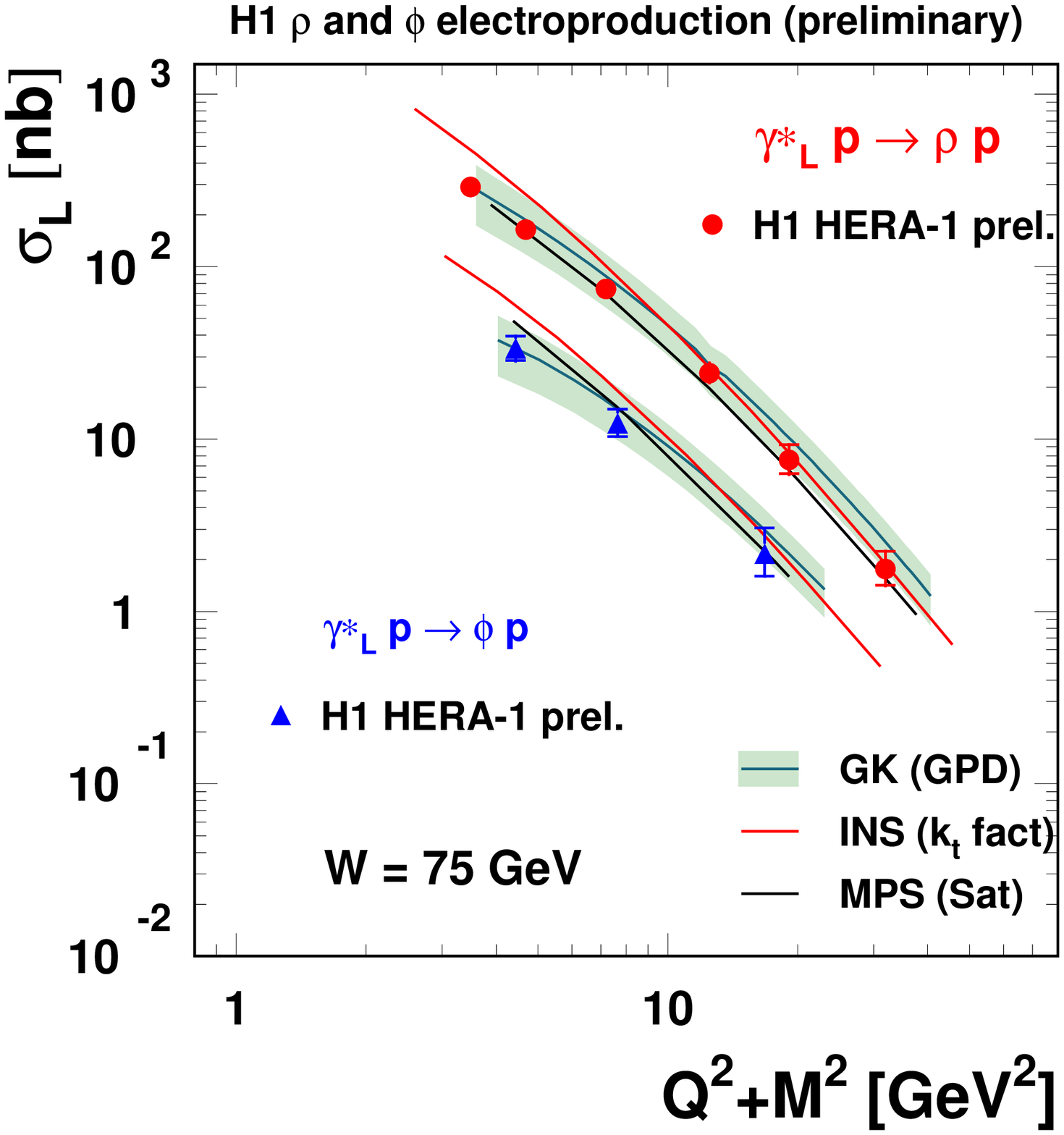}
  \includegraphics[width=0.37\columnwidth]{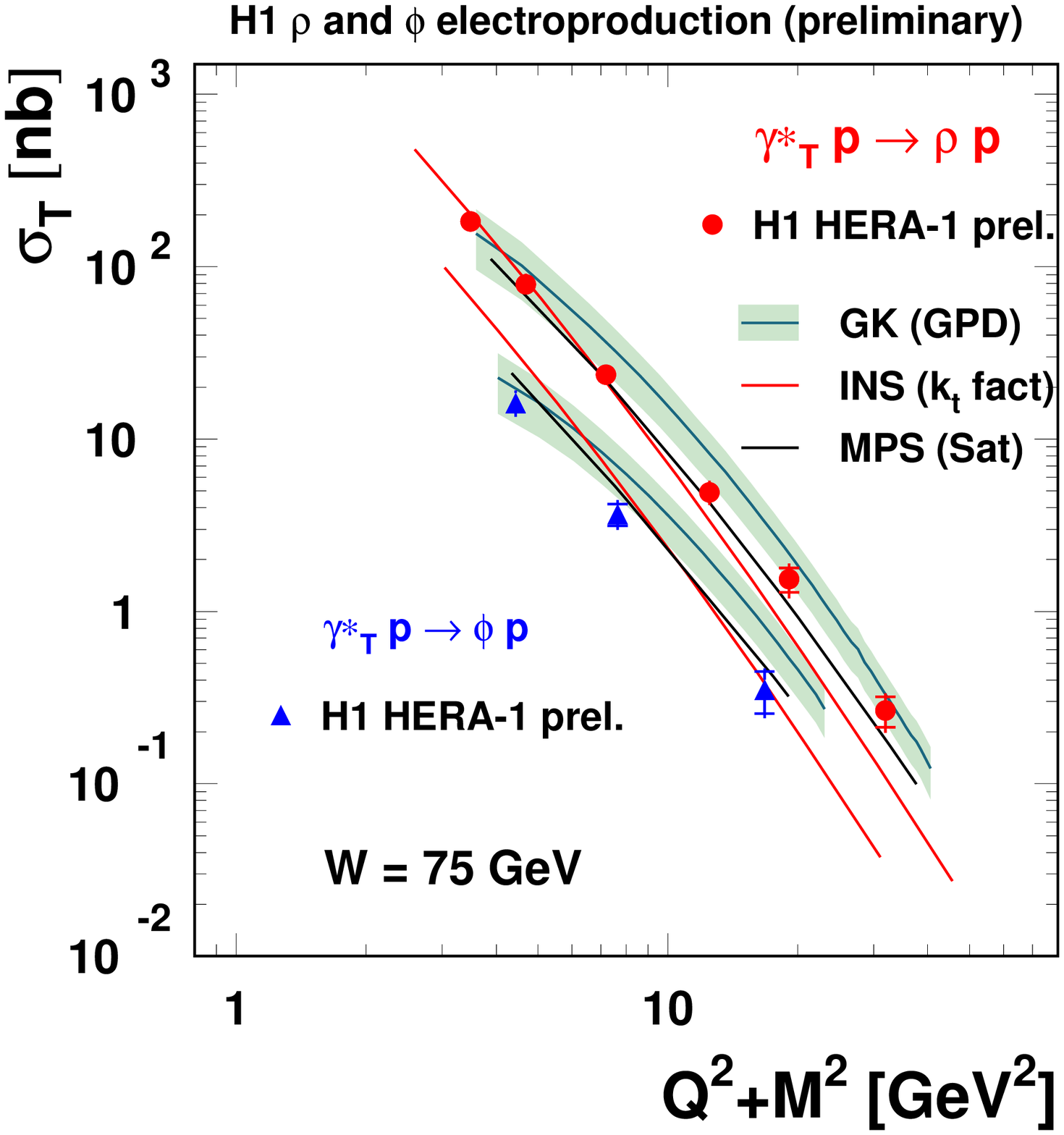}
\caption{The $\rho$ and $\phi$ cross sections induced by the longitudinally ($\sigma_L$) and transversely ($\sigma_T$) polarised photons as a function of $Q^2+M^2$, compared to theoretical predictions.}
\label{fig4}
\end{figure}

Figure \ref{fig4} shows the H1 measurement of $\rho$ and $\phi$ cross sections induced by the longitudinally ($\sigma_L$) and transversely ($\sigma_T$) polarised photons as a function of $Q^2+M^2$ \cite{rhophih1}. The results are compared to the theoretical models based on pQCD \cite{h1model1,h1model2,h1model3}. The models describe the $\sigma_L$ dependence well, but are not able to reproduce the $\sigma_T$ cross section. The ZEUS measurements of the $\rho$ cross sections \cite{rhozeus} were also compared to the theoretical models, which differ in the assumptions on the vector meson wave function and in the modelling of the dipole-proton interaction \cite{zeusmodel1,zeusmodel2,zeusmodel3}. None of the models is able to reproduce the $\rho$ cross section in the entire kinematic range of the measurement. The data are sensitive to the different parametrisations of PDFs and therefore have the potential to provide constraints on PDFs. 

The dependence of the cross sections on generalised gluon distributions (GPDs) was studied by the H1 experiment using the DVCS data \cite{dvcsh1}. The DVCS results were compared with the pQCD predictions in terms of two dimensionless observables, $S(Q^2)$ and $R(Q^2)$, which measure the $Q^2$ evolution of the GPDs and the ratio of the GPDs to PDFs (ie. skewing effect), respectively. Both variables are well described by the theoretical model based on NLO QCD calculations \cite{gpd}
. Furthermore, for the first time in a collider mode, the contribution from the real part of the DVCS amplitude was observed. The sensitivity to this contribution arises from the interference term between the DVCS and the QED Bethe-Heitler processes, which changes the sign with the charge of the lepton beam. The beam-charge asymmetry (BCA) was measured as a function of $\phi$, the angle between the two planes defined by the incoming and outgoing electron and the virtual photon and the outgoing proton \cite{beamash1}. The BCA was fitted with the form $p_{1}cos\phi$ and the non-zero value of parameter $p_1 = 0.17 \pm 0.03 \pm 0.05 (syst.)$ was extracted, providing additional information on gluon GPDs.

\vspace{-0.2cm}

\section{Summary}

The recent high-precision measurements of light vector mesons and DVCS have been reported. The observed features of the data are consistent with the expectations from pQCD. The DVCS data are well described by the theoretical models, but none of the models is able to describe all the features of the data for light vector mesons.



\begin{thebibliography}{9}   

\bibitem{dvcstheo1} L.L Frankfurt, A. Freund, M. Strickman {\em Phys. Rev.} {\bf D58} (1998) 114001
\bibitem{rhozeus} ZEUS Coll., S. Chekanov et al., {\em PMC Physics} {\bf A~1} (2007) 6
\bibitem{rhophih1} H1 Coll., A. Aktas et al., H1prelim-08-013, sub. to DIS08, London.
\bibitem{upsilonzeus} ZEUS Coll., S. Chekanov et al., ZEUS-prel-07-015, sub. to EPS07, Manchester.
\bibitem{dvcszeus} ZEUS Coll., S. Chekanov et al., ZEUS-prel-07-016, sub. to EPS07, Manchester.
\bibitem{schilling} K. Schilling, G. Wolf, {\em Nucl. Phys.} {\bf B61} (1973) 381
\bibitem{h1model1} I.P. Ivanov, N.N. Nikolaev, A. A. Savin {\em Phys. Part. Nucl.} {\bf 37} (2006) 1
\bibitem{h1model2} S. V. Goloskokov, P. Kroll {\em arXiv:hep-ph/0708.3569} (2007)
\bibitem{h1model3} C. Marquet, R. Peschanski, G. Soyez {\em Phys. Rev.} {\bf D76} (2007) 034011
\bibitem{zeusmodel1} L. Frankfurt, W. Koepf, M. Strikman {\em Phys. Rev} {\bf D 57} (1998) 512
\bibitem{zeusmodel2} A.D. Martin, M.G. Ryskin, T. Teubner {\em Phys. Rev.} {\bf D 62} (2000) 014022
\bibitem{zeusmodel3} H. Kowalski, L. Motyka, G. Watt {\em Phys. Rev.} {\bf D 74} (2006) 074016
\bibitem{dvcsh1} H1 Coll., A. Aktas et al., {\em Phys. Lett.} {\bf B659} (2008) 796
\bibitem{gpd} A. Freund, {\em Phys. Rev.} {\bf D68} (2003) 096006
\bibitem{beamash1} H1 Coll., A. Aktas et al., H1prelim-07-011, sub. to DIS07, Munich.

\end{thebibliography}
\end{document}